\documentstyle[12pt]{article}
\begin{document}
\rightline{RU9248}
\vspace*{.5in}
\begin{center}
Generalized $^3P_0$ and $^3S_1$ Annihilation Potentials for $\bar{p}p$ Decay\\
into Two Mesons based on a Simple Quark Model\\

\vspace*{.8in}

George Bathas and W.M. Kloet\\

\vspace*{.5in}

Department of Physics and Astronomy, Rutgers University\\
P.O. Box 849 Piscataway NJ 08854\\

\end{center}
\vspace{1.2in}

\centerline{Abstract}

\bigskip

\indent Within the quark model a generalization is proposed of the
commonly used
annihilation potential to describe antiproton-proton annihilation into two
mesons, the so-called $^3P_0$ and $^3S_1$  mechanisms. This
generalized potential
treats the two mechanisms in a more symmetric way, has additional angular
dependence, and results in an expanded set of selection rules.

\newpage

\baselineskip=18pt

\noindent 1. Introduction\\

     As a first step on the way to find a microscopic model for the
annihilation of antiproton and proton into two mesons, the process is often
described in the simple quark model by diagrams representing the so-called
vacuum ($^3P_0$) and one-gluon ($^3S_1$) mechanisms. A detailed
description appears in
a recent review of $\bar{p}p$ annihilation [1]. In particular the  $^3P_0$
mechanism was
inspired by earlier work on meson decay [2]. Since several years many authors
have studied the $\bar{p}p$ annihilation process, prefering just one particular
mechanism or some combination of $\bar{q}q$ annihilation into vacuum
and one-gluon. We
mention refs.[3-21], but a more complete list of references can be found in
the review of ref.[1]. In general the annihilation is described by a potential

\begin{eqnarray}
V_{\rm ann} ~= ~ V(^3P_0) ~+~ \lambda ~ V(^3S_1),
\end{eqnarray}

\noindent where $\lambda$  is the relative strength of the two mechanisms. In
the  $^3P_0$  mechanism
the $\bar{q}q$ pair annihilates into the effective ``vacuum'' state
$J^\pi  = 0^+$.  In the $^3S_1$
mechanism the $\bar{q}q$ pair annihilates into the effective
``one-gluon'' state $J^\pi = 1^-$.  Examples of quark diagrams for the
$\bar{p}p$ annihilation into two mesons are
shown in fig.1. Because quark diagrams with real gluon exchanges are
infinitely more complex, the above mechanisms should be regarded as the first
two leading operators in an expansion of the annihilation mechanism of
the $\bar{q}q$
system into terms of increasing $J^\pi$. Since the exchanged ``vacuum''
and ``one-gluon'' are only effective this means in particular that in both
cases
transfer of momentum occurs , a fact that is neglected in previous
descriptions  of the  $^3P_0$  mechanism. As a result previous
treatments of  $^3P_0$  and
$^3S_1$ are somewhat asymmetric. Most versions of the vacuum term only
contribute
to $\ell_{\rm MM}$ =0 and $\ell_{\rm pp}$ = 1, while the one-gluon
term contributes to an entire
series of angular momentum states of the $\bar{p}p$ system.\\

\noindent 2. Generalized potential\\

     In this Letter we propose to describe the annihilation mechanism
of $\bar{p}p$
into two mesons by the potential  $V_{\rm ann} = V(^3P_0) + \lambda~
V(^3S_1)$, but allowing
momentum transfer in both mechanisms. The relative strength $\lambda$
is treated as a
parameter. Additional diagrams where two $\bar{q}q$ pairs are
annihilated and another
$\bar{q}q$ pair is created, should also be considered in further
generalizations, in
particular when annihilation into strange mesons occurs (see fig.1(c)). This
can even lead to additional $J^\pi$ terms in the expansion of the annihilation
operator, but they are ignored at this time.

     The nucleon (and antinucleon) wave function is described as a Gaussian

\begin{eqnarray}
\Psi_N (\vec{r}_1, \vec{r}_2, \vec{r}_3) ~=~ N_N~ {\rm exp}~[~ -
\frac{\alpha}{2} \Sigma (\vec{r}_i - \vec{r}_N)^2~]~ X_N ~
 {\rm (spin,~isospin,~color)},
\end{eqnarray}

\noindent where $\vec{r}_i$ are the quark coordinates and $\vec{r}_N$ is the
nucleon coordinate. An S-wave meson wavefunction is given by

\begin{eqnarray}
\Phi_M (\vec{r}_1, \vec{r}_4) ~ = ~ N_M ~ {\rm exp}~[~
-~\frac{\beta}{2} \Sigma (\vec{r}_i - \vec{r}_M)^2] ~ X_M
 {\rm (spin,~isospin,~color)},
\end{eqnarray}

\noindent where $\vec{r}_1$ and $\vec{r}_4$ are respectively the quark and
antiquark coordinates, and $\vec{r}_M$
is the coordinate of the meson. Typical parameter values are $\alpha$
= 2.8 fm$^{-2}$ and
$\beta$ = 3.23 fm$^{-2}$, giving a nucleon radius of 0.60 fm and a
meson radius of 0.48 fm.

      The $\bar{p}p$ annihilation is then described by a non-local transition
potential in terms of the relative $\bar{N}N$ coordinate $\vec{r}$,
the relative two-meson
coordinate $\vec{r}~'$, and the spin-operator $\vec{\sigma}$, obtained
by integrating out the unconstrained coordinates of the quarks and antiquarks.

      As an example we consider from here on explicitly the reaction
$\bar{p}p \rightarrow \pi^- \pi^+$, using only the diagrams 1(a) and
1(b). Here it is assumed that the pion can
be described by a quark wave function of the form of eq.(3).\\

\noindent 3. Vacuum Mechanism\\

      For the ``vacuum'' mechanism one obtains from a single diagram a
potential

\begin{eqnarray}
V_{\rm single}(^3P_0) (\vec{r}~',\vec{r}~)& \sim& \nonumber \\
\{ A_V~ i\vec{\sigma}.\vec{r} ~' ~& +& ~ B_V ~i \vec{\sigma}.\vec{r}~\}~
{\rm exp}( A r~'^2 + B r^2  + C \vec{r} ~'.\vec{r}~),
\end{eqnarray}

\noindent where $A_V$, $B_V$, A, B, and C are functions of the parameters
$\alpha$ and $\beta$. Summing
over all diagrams in case of the ``vacuum'' term one obtains the form

\begin{eqnarray}
V_{\rm total}(^3P_0)(\vec{r} ~',\vec{r}~) &~\sim &\\ \nonumber
\{ A_V ~ i\vec{\sigma}.\vec{r} ~'~  {\rm sinh}(C\vec{r} ~'.\vec{r}~)~ &+&~ B_V~
i\vec{\sigma}.\vec{r} ~ {\rm cosh}(C\vec{r} ~'.\vec{r}~)\} ~{\rm exp}
(Ar~'^2 ~+~  B r^2),
\end{eqnarray}

\begin{eqnarray}
{\rm where} ~  A_V~  =~ \frac{\alpha ( \alpha + \beta)}{ 2 (4 \alpha +
3 \beta)},
\end{eqnarray}

\begin{eqnarray}
B_V ~=~ \frac{ 3 ( 5 \alpha^2 + 8 \alpha \beta + 3 \beta^2)}{2 (4
\alpha + 3 \beta)} ,
\end{eqnarray}

\begin{eqnarray}
A ~=~ - ~ \frac{\alpha (5 \alpha + 4 \beta)}{2 (4 \alpha + 3 \beta)} ,
\end{eqnarray}

\begin{eqnarray}
B ~=~ -  ~ \frac{ 3 ( 7 \alpha^2 + 18 \alpha \beta + 9 \beta^2)}{8(4
\alpha + 3 \beta)} ,
\end{eqnarray}

\begin{eqnarray}
C ~=~ - ~ \frac{3 \alpha ( \alpha + \beta) }{2 (4 \alpha + 3 \beta) }.
\end{eqnarray}

      The potential V($^3P_0$) ($\vec{r}~'$,$\vec{r}$) is even in
$\vec{r}~'$ and odd in $\vec{r}$. It contributes
to $\ell_{\pi \pi}$ = 0, 2, 4, $\ldots$ , and $\ell_{pp}$ = 1, 3, 5,
$\ldots$ . In other words V($^3P_0$) acts in
$J^\pi$ = 0$^+$, 2$^+$, 4$^+$, $\ldots$ waves (e.g.  $^3P_0$,  $^3P_2$
- $^3F_2$, $^3F_4$ - $^3H_4$, $\ldots$ ) with isospin I = 0.

     The present potential V($^3P_0$) of eq.(5), where the linear
momentum of the
annihilating $\bar{q}q$ pair is transfered to one of the final quarks
or antiquarks,
has to be compared with the standard  V($^3P_0$) expression, where no linear
momentum is transfered. The standard potential is

\begin{eqnarray}
V'~ (^3P_0) (\vec{r} ~',\vec{r}) ~ \sim~  B_V' ~i \vec{\sigma}.\vec{r}
{}~{\rm exp}(A' r'^2 ~ + ~B' r^2),
\end{eqnarray}

\noindent where $B_V'$ = $\alpha$ + $\beta$, A$'$ = $-$ $\alpha$/2, and
B$'$ = $-$ 3 ( $\alpha$ + 3$\beta$)/8. The
expression of eq.(5) has additional angular dependence due to the presence of
the C$\vec{r}~'$.$\vec{r}$ term in the exponent and the additional
$A_V$ i$\vec{\sigma}$.$\vec{r}~'$ term. In previous
cases (without momentum transfer) only a transition between $\ell_{\pi
\pi}$=0 and $\ell_{pp}$=1
was possible (e.g. $^3P_0$) , while with the present form all even J values
contribute.\\

\noindent 4. One-Gluon Mechanism\\

      The ``one-gluon'' exchange term splits in a transversal part and a
longitudinal part. The transversal part becomes

\begin{eqnarray}
V_T (^3S_1)~ (\vec{r} ~',\vec{r})& ~\sim & \nonumber \\
\{A_T~ i\vec{\sigma}.\vec{r} ~'~ {\rm cosh}(C\vec{r} ~'.\vec{r}) & + & B_T~
i\vec{\sigma}.\vec{r} ~{\rm sinh}(C\vec{r} ~'.\vec{r})\}~{\rm exp} (A
r'^2  + B r^2),
\end{eqnarray}

\noindent where A, B, and C are the same as in eqs.(8-10) , while

\begin{eqnarray}
A_T ~ = ~ -~ \frac {2 \alpha ( \alpha + \beta)}{4 \alpha + 3 \beta},
\end{eqnarray}

\begin{eqnarray}
B_T ~=~ - ~ \frac{3 \alpha ( \alpha + \beta)}{ 4 \alpha + 3 \beta}.
\end{eqnarray}

      The potential $V_T$($^3S_1$)($\vec{r}~'$,$\vec{r}$) is odd in
$\vec{r}~'$ and even in $\vec{r}$, and contributes
to $\ell_{\pi \pi}$ = 1, 3, 5, ${\ldots}$ , and $\ell_{pp}$ = 0, 2, 4,
${\ldots}$ . Therefore $V_T$($^3S_1$) acts in $\bar{p}p$
states with  J$^\pi$ = 1$^-$, 3$^-$, 5$^-$, ${\ldots}$ waves  (e.g.
$^3S_1 - ^3D_1$, $^3D_3 - ^3G_3$, $^3G_5 - ^3I_5$, ${\ldots}$)
with isospin I = 1. The basic form of $V_T$($^3S_1$) is the same as
used before in the literature and is described for example in ref.[5].

     The second ``one-gluon'' term is longitudinal and can be written as

\begin{eqnarray}
V_L(^3S_1)~~(\vec{r} ~',\vec{r}) ~& \sim& \\ \nonumber
\{ A_L ~  i\vec{\sigma}.\vec{r} ~'~ {\rm sinh}(C\vec{r} ~'.\vec{r}) &+& B_L~
i\vec{\sigma}.\vec{r}~ {\rm cosh}(C\vec{r} ~'.\vec{r})\}~ {\rm exp}( A
r'^2  + B r^2),
\end{eqnarray}

\begin{eqnarray}
{\rm where} ~  A_L ~ = ~ - ~ \frac{\alpha ( \alpha + \beta)}{ 4 \alpha
+ 3 \beta} ,
\end{eqnarray}

\begin{eqnarray}
B_L ~=~ \frac{9 ( \alpha + \beta) ^2}{2 ( 4 \alpha + 3 \beta)}
\end{eqnarray}

\noindent $V_L$($^3S_1$) has the same symmetry as V($^3P_0$) of eq.(5)
and therefore acts in the
same J$^\pi$ waves. The parameters A, B, and C in eq.(15) are again
the same as
eqs.(8-10).\\

\noindent 5. Concluding Remarks\\

     The generalized potential  $V_{\rm ann} = V(^3P_0) + \lambda
V(^3S_1)$  is a function of the
relative two-meson coordinate $\vec{r}~'$, the relative $\bar{p}p$
coordinate $\vec{r}$, and the spin-operator $\vec{\sigma}$. All
coefficients are functions of the nucleon and meson size-parameters $\alpha$
and $\beta$. The various parts of $V_{\rm ann}$ satisfy systematic selection
rules. Both vacuum and one-gluon exchange mechanisms are treated on equal
footing in a symmetric way. Both can contribute in a series of
J$^{\pi}$ channels.

      From the point of view of the range of this potential in either variable
$\vec{r}$ or $\vec{r}~'$, it is interesting to mention typical values
of  A, B, and C. With the
values adopted above for $\alpha$ and $\beta$ one finds A = -1.80
fm$^{-2}$, B = -5.59 fm$^{-2}$, and
C = - 1.21 fm$^{-2}$, while for the potential V$'$ ($^3P_0$) of eq.(11)
(where there is
no momentum transfered), the corresponding values are A$'$ = -1.40
fm$^{-2}$, B$'$ = - 4.68 fm$^{-2}$, and C$'$ = 0.

     The above approach can be extended to include further diagrams (e.g.
fig.1(c)). It can be applied to $\bar{p}p$ annihilation into other
mesons, including
strange mesons or into strange baryons such as $\bar{\Lambda}
\Lambda$. It may also be interesting
to see whether these generalized  $^3P_0$  and $^3S_1$ potentials
alter  predictions of
the  branching ratios for the various $\bar{p}p$ annihilations.\\

\noindent Acknowledgement\\

     We wish to acknowledge stimulating discussions with E.M. Henley and M.A.
Alberg. We would also like to thank L. Zamick for helpful comments on the
manuscript.\\

\noindent References\\

\noindent [1]  C.B. Dover, T. Gutsche, M. Maruyama, A. Faessler, Prog. in
Particle and
Nuclear Physics 29 (1992) 87.\\
\noindent [2]  A. Le Yaouanc, L. Oliver, O. Pene, J.-C. Raynal, Phys.Rev. D8
(1973)
2223.\\
\noindent [3]  A.M. Green, J.A. Niskanen, Nucl.Phys. A412 (1984) 448; A430
(1984) 605;
Mod.Phys.Lett. A1 (1986) 441.\\
\noindent [4]  C.B. Dover, P.M. Fishbane, Nucl.Phys. B244 (1984) 349.\\
\noindent [5]  A.M. Green, J.A. Niskanen, S. Wycech, Phys.Lett. 139B (1984) 15;
172B
(1986) 171.\\
\noindent [6]  S. Furui, A. Faessler, S. Khadkikar, Nucl.Phys. A424 (1984)
495.\\
\noindent [7]  M. Maruyama, T. Ueda, Progr.Theor.Phys. 73 (1985) 1211; 78
(1987) 841.\\
\noindent [8] J.A. Niskanen, F. Myhrer, Phys.Lett. 157B (1985) 247.\\
\noindent [9]  J.A. Niskanen, V. Kuikka, A.M. Green, Nucl.Phys. A443
(1985) 691.\\
\noindent [10]  A.M. Green, V. Kuikka, J.A. Niskanen, Nucl.Phys. A446
(1985) 543. \\
\noindent [11] M Kohno and W. Weise, Phys.Lett. 152B (1985) 303; Nucl.Phys.
A454 (1986)
429.\\
\noindent [12]  E.M. Henley, T. Oka, J. Vergados, Phys.Lett. 166B (1986) 274.\\
\noindent [13]  S. Furui, Z.Phys. A325 (1986) 375.\\
\noindent [14]  C.B. Dover, P.M. Fishbane, S. Furui, Phys.Rev.Lett. 57 (1986)
1538.\\
\noindent [15] A.M. Green, G.Q. Liu, Z.Phys. A331 (1988) 197.\\
\noindent [16]  U. Hartman, E. Klempt, J. Korner, Z.Phys. A331 (1988) 217.\\
\noindent [17] T. Gutsche, M. Maruyama, A. Faessler, Nucl.Phys. A503
(1989) 737.\\
\noindent [18]  H. Genz, M. Martinis, S. Tatur, Z.Phys. A335 (1990) 87.\\
\noindent [19]  L. Mandrup, A.S. Jensen, A. Miranda, G.C. Oades, Phys.Lett.
B270 (1991)
11.\\
\noindent [20]  M.A. Alberg, E.M. Henley, L. Wilets, Z. Phys. A331
(1988) 207.\\
\noindent [21]  M.A. Alberg, E.M. Henley, L. Wilets, P.D. Kunz, Nucl.Phys. A508
(1990)
323c.\\

\noindent FIGURE CAPTIONS\\

\noindent FIG. 1. Quark diagrams corresponding to $\bar{p}p$
annihilation into two mesons.

\end{document}